\documentclass[pdflatex,sn-mathphys]{sn-jnl}

\usepackage{subfigure}
\usepackage{multicol}

\jyear{2023}%

\begin{document}

\title[Multi-Platform Bot Detector]{Assembling a Multi-Platform Ensemble Social Bot Detector with Applications to US 2020 Elections}


\author*[1]{\fnm{Lynnette Hui Xian} \sur{Ng}}\email{lynnetteng@cmu.edu}

\author[1]{\fnm{Kathleen M.} \sur{Carley}}\email{carley@andrew.cmu.edu}

\affil[1]{\orgdiv{Center for Computational Analysis of Social and Organizational Systems}, \orgname{Carnegie Mellon University}, \orgaddress{\street{4665 Forbes Avenue}, \city{Pittsburgh}, \postcode{15213}, \state{PA}, \country{USA}}}


\abstract{
Bots have been in the spotlight for many social media studies, for they have been observed to be participating in the manipulation of information and opinions on social media. These studies analyzed the activity and influence of bots in a variety of contexts: elections, protests, health communication and so forth. Prior to this analyses is the identification of bot accounts to segregate the class of social media users. In this work, we propose an ensemble method for bot detection, designing a multi-platform bot detection architecture to handle several problems along the bot detection pipeline: incomplete data input, minimal feature engineering, optimized classifiers for each data field, and also eliminate the need for a threshold value for classification determination.  With these design decisions, we generalize our bot detection framework across Twitter, Reddit and Instagram. We also perform feature importance analysis, observing that the entropy of names and number of interactions (retweets/shares) are important factors in bot determination. Finally, we apply our multi-platform bot detector to the US 2020 presidential elections to identify and analyze bot activity across multiple social media platforms, showcasing the difference in online discourse of bots from different platforms.
}

\keywords{bot detection, Twitter, Reddit, Instagram, social media, interpretability, machine learning, US 2020 elections}

\maketitle

\section{Introduction}
Social media bots, which are automated accounts, have been shown to participate in election interference \cite{ferrara2016rise}, opinion manipulation in vaccination efforts \cite{ng2022pro} and even extremism campaigns \cite{ferrara2016predicting}. The field of social cybersecurity is concerned with the problem of identifying these bot accounts because the bot campaigns can lead to negative offline impacts like protests.

A suite of bot detection models have been developed to characterize users on social media space as bot or humans. These bot detection models use techniques from feature-based detection to temporal detection to graph-based detection. However, the training and inference of these bot detection models often involves huge feature spaces, i.e. 1000+ extracted user features \cite{yang2019arming}; or extensive data collection, i.e. temporal methods require longitudinal data and graph-based methods require network data. While the increase in feature space often results in improved performance \cite{yang2019arming}, data collection becomes harder. With data collection requirements come the issue of incomplete data input: missing fields in input data due to data collection limitations, change in data formats, or unavailability of field. Unfortunately, the prevailing models typically rely on the completeness of account data to make a prediction. This is because models are typically tuned by the union of data features, and are thus unable to make a prediction with incomplete data.

After the data input is passed through bot detection algorithms, a bot probability score is typically returned. This score is between 0 and 1 and indicates the likelihood of an account being a bot. A threshold value is usually defined, where if the score is above the threshold, the account is deemed as a bot; and as a human otherwise. However, the threshold is usually arbitrarily determined, and values used ranged from 0.2 to 0.7, leading to a false positive problem \cite{rauchfleisch2020false,ng2022stabilizing,yang2019arming}. In fact, for Elon Musk's bot estimate in July 2022 during his Twitter acquisition negotiations, one key question was: what was the threshold value Musk used? \cite{clayton_2022} The choice of a threshold value can affect the determination of the proportion of bots, which will be different should different analysts choose different thresholds. 

In this paper, we address the aforementioned problems, by designing a multi-platform ensemble architecture. Our architecture uses a small set of features for bot detection, separating the features into data chunks that represent user, user metadata and content features. Not only does this enable fine-tuning of separate classifiers for each data field, it also handles the problem of incomplete data where prediction can be made with the remaining classifiers. We aggregate bot/human probabilities before taking the larger value, eliminating the need to determine a threshold value.

As a result of these design decisions, we are able to generalize our bot detection framework across multiple platforms: Twitter, Reddit, and Instagram. Many of the bot detection models are currently constructed for the Twitter platform and there are few that analyze bot activities on other social media platforms, much less multiple platforms within a single bot detection architecture. We leveraged on training separate models for each data field in a parallel fashion before combining data across platforms. In this paper we also aim to improve the running capability of bot detection classifiers. Therefore, we leverage mostly on simpler tree-based classifiers instead of focusing on deep-learning based or graph-based classifiers, with the intent that our bot detection classifier can be run on a variety of machines, from low-powered to high-powered machines, thereby facilitating the analysis and research of bot detection.
Our tree-based classification ensemble runs extremely quickly, completing 759 users in 3.9 minutes on an Intel Xeon-1250 CPU, which can facilitate large-scale bot detection. Across a series of 7 Twitter, 1 Reddit and 1 Instagram datasets, we show that our model outperforms baselines with an average accuracy of 75.47\%. 

The layout of this paper is as follows: in \autoref{sec:litreview} we provide a brief literature review of bot detection models and bot detection on multiple social media platforms. Then, in \autoref{sec:methodology} we describe the construction of the bot detection model. After building our bot detection model, we applied it to a slice of the online discourse on two social media platforms in \autoref{sec:application}, illustrating the use of our bot detection model on multiple platforms. Finally, we discuss the observations in our paper in \autoref{sec:discussion} and provide concluding remarks in the final section.

\section{Literature Review}
\label{sec:litreview}
Social media bots, or fondly called ``bots", refer to social media users that are software-controlled and can automatically perform a series of tasks. These types of user accounts are of keen interest to the social cybersecurity community because they have been observed to perform malicious activities online, which can affect the peace of society. They have observed to be used to infiltrate political discourse and spread misinformation. During the 2010 US midterm elections, social bots were already observed to have been flooding the social media space with their support for some candidates and smear their opponents by injecting thousands of tweets pointing to websites with fake news \cite{ferrara2016rise}. Bots are also used by countries for digital diplomacy, to put forth a desired narrative facing the online public \cite{ng2023deflating,jacobs2023tracking}.
Bots working together in a coordinated fashion have also been known to apply social pressure on to other users, causing them to change their opinion towards key topics. This was observed in the case of the 2021 coronavirus vaccination debate, where users surrounded by coordinating active bots change their stance towards the vaccine, potentially resulting in an anti-vaccine stance and the real-world refusal of the vaccine \cite{ng2022pro}.

A suite of bot detection algorithms have been developed for the detection of automated social media bot accounts in key events such as elections and protests. A bot detection algorithm classification is a binary classification task: classifying whether a user is a bot or a human. For such a task, there are essentially two main approaches: a supervised learning approach where data labels are known and the model is trained on the segregation of data labels, and an unsupervised learning approach where the model discovers hidden patterns within the dataset.

Supervised learning models work based on identifying distinct set of features for each class in a dataset labeled bot and humans. These detection algorithms can be grouped into three types: feature-based, temporal-based and graph-based algorithms. All three types of algorithm features can be combined to be fed into a machine learning classifier, as in the case of T-Bot, which uses profile-based, user activity based and social network based features in its classification \cite{gera2022t}. Feature-based algorithms are algorithms that apply machine learning algorithms to features engineered from user and content information \cite{yang2019arming,yang2020scalable}. Examples of such features are: average number of hashtags used per post, number of URLs used per post, average number of punctuations used per post, number of interactions per post (i.e. retweet, quote tweet, shares, likes), sentiment of the post and so forth. The machine learning models built on features range from logistic regression classifiers \cite{heidari2021empirical,kantepe2017preprocessing}, to support vector machines \cite{pratama2019social}, to neural network-based classifiers \cite{kudugunta2018deep}.

Temporal-based models characterize accounts through time series pattern analysis and behavior activity occurrence \cite{cresci2018fake,mazza2019rtbust}. Another strategy is to make use of the patterns of inter-arrival times between posts and extract features to represent the circadian rhythm and cultural and environmental influences of a user for use in the classification model \cite{cai2017behavior}. The time interval between posts can also be processed to derive parameters that characterize the burst patterns or information entropy of posts and use them as classification features \cite{wu2021novel}. These temporal features that are derived are eventually fed into a machine learning model which differentiates whether the user is a bot or human. These time-series methods, however, require a good length of post data across time of each account, which can be difficult to acquire given the volume of accounts and the platform's rate limits.

Graph-based models which make use of an account's social network graph to enhance predictions with information inferred from the account's neighbors \cite{feng2021twibot}. This technique builds on the concept of homophily, that users tend to interact with other similar users. The technique thus makes use of a matrix that reveals the connections formed between users, assuming the connections are formed with confidence, i.e. the same type of users tend to form connections with each other. This matrix is then put through a machine learning model, for example a graph-based regression model, to differentiate the user classes. Since a graph-based approach constructs matrices based on connections between users, it can be extended for use across many social media platforms \cite{al2018prediction}. One drawback of graph-based models, however, is that while they can be fairly accurate in determining bot-likelihood from an account's friends ($\sim$85\% accuracy \cite{feng2021twibot}), collecting the other users that a user is following/ follows him can be time and resource intensive. Graph-based methods also works mostly on Twitter and Instagram data for those platforms do have a follow/following feature, but Reddit does not reveal the users that follow a user.

Unsupervised learning approaches use anomaly detection or time-series based methods to extract connectivity of suspicious accounts. DeBot \cite{chavoshi2016debot} is an unsupervised classification algorithm that makes use of temporal patterns to determine the presence of bot accounts, inferring the presence of bot accounts using time series spikes. BotWalk compares each new user to a seed of bot/human user using an ensemble anomaly detection method \cite{minnich2017botwalk}. Time-series based methods include algorithms like MulBot and RTBust. MulBot infers bot accounts through multivariate time series statistics of the user posts as features \cite{mannocci2022mulbot}. RTBust constructs a univariate time series based on the time difference between retweets, which then is fed into an LSTM autoencoder \cite{mazza2019rtbust}.

Finally, ensemble-based classification models are approaches that combine multiple classification models together to increase the accuracy of differentiating a user. \cite{sayyadiharikandeh2020detection} developed an Ensemble of Specialized Classifiers to detect different types of bots, like spam bots and fake follower bots. This ensemble is made up of multiple Random Forests classifiers and aggregated through a voting system. Similarly, \cite{dimitriadis2021social} trained Random Forest classifiers based on content, user, temporal and social network features to differentiate between different bot types (e.g., political bots, spam bots, social bots etc). The value of an ensemble-based approach is that it can work well to produce outputs for multiple binary classification tasks that have disparate outcomes and inputs, then aggregate them together for the final outcome.

Most of the bot detection algorithms are designed for the detection of bots on Twitter, a microblogging platform. One of the commonly used Twitter bot detection algorithm is Botometer, which uses over 1000 features extracted from social media profiles to perform its classification. It uses the Twitter API to query the platform live, in order to provide the latest update as to the bot likelihood of the account, but in doing so is unable to perform predictions for historical data \cite{yang2020scalable}. 

Reddit is a forum-like site which is organized by interests, termed subreddits. Temporal analysis methods have been used in Reddit bot detection to classify accounts in terms of their bot-likelihood based on their temporal bursts between comments and their network connectivity between subreddits \cite{hurtado2019bot}. Another feature engineering method analyzes the presence of a user account making Reddit submissions with same titles and the comment activity to characterize bot activity \cite{saeed2022trollmagnifier}.

In terms of bots in the image-based social media platform Instagram, classification algorithms have been developed with logistic regression, naive bayes or support vector machines that take in profile features such as follower/following counts, number of digits in the account username and so forth in order to provide a bot/human differentiation \cite{akyon2019instagram}. Instagram bot accounts that impersonate politicians, news agencies and sports stars have also been differentiated through clustering algorithms that makes use of profile metrics like number of posts, comments, likes and so forth \cite{zarei2019typification}. 

\section{Methodology}
\label{sec:methodology}
This section specifies the building of the multi-platform social bot detector, beginning with the description of the training datasets, then the description of the machine learning algorithms used in the construction and evaluation of the bot detection model, and finally we perform an evaluation on an external dataset.

\subsection{Data}
In building our multi-platform social bot detector, we used datasets from Twitter, Reddit and Instagram. By using datasets across multiple social media platforms, and thereafter training the bot detection model on this aggregated dataset, we are able to build a bot detection model that is able to analyze multiple social media platforms. 

We used the following datasets: (1) Seven Twitter datasets extracted from the OSOME bot repository\footnote{\url{https://botometer.osome.iu.edu/bot-repository/datasets.html}}. These datasets contain only user information from Twitter profiles and we rehydrated them with the Twitter V2 API in June 2022. Some accounts have since been suspended prior to the rehydration and only partial user and content information are available. (2) Reddit dataset was self-curated through extracting the top 500 ``bad bots" flagged by Reddit users on B0tRank\footnote{\url{https://botrank.pastimes.eu}}. The dataset is enhanced with the human users who reported the bots. (3) Instagram bot dataset was self-curated through a purchase of fake follower bots. These bots follow a public account to increase the number of followers, providing the illusion of account popularity, increasing the influence of the account.
We then harmonized the naming conventions data fields of all three platforms. Twitter provides the most data fields, whereas Reddit and Instagram do not posses all the fields, hence we work with partial data. 

For each of the social media accounts, we identified several fields that are important to be used in our bot detection model: user name, screen name, description, posts and user metadata (i.e., number of followers, number of following). However, not all datasets provided all the information. This is due to two reasons: profile suspension on Twitter or that the platform does not provide the information. In these cases, we identify these fields as partially available for the datasets. For it to be useful, our social bot detector needs to be able to handle datasets in which the fields are partially available, and make use of the available fields to make a best-guess decision on whether the account is a bot or not.

\autoref{tab:dataset_statistics} summarizes the dataset bot/human composition and data field availability (present, not present, partially present). We use an aggregation of these datasets to construct a bot detection algorithm. Many of these datasets contain incomplete information about the users, due to the unavailability of data at collection time. Therefore, our bot detection algorithm needs to be able to provide its best-guess prediction under the circumstances of incomplete data. We also have datasets from three different social media platforms, and we aim to construct a bot detection algorithm that is generic enough to apply to all three platforms. In our work, data was only collected from public accounts and no attempt was made to access or use information that was not publicly available from the social media sites.

\begin{table}[!ht]
\centering
\begin{tabular}{|p{2.5cm}|p{1.5cm}|p{1.5cm}|p{1cm}|p{1.5cm}|p{1cm}|p{1.5cm}|}
\hline
\textbf{Dataset} & \textbf{Users (\% Bots)} & \multicolumn{5}{|c|}{\textbf{Data Present}} \\ \hline
& & User name & Screen Name & Description & Posts & User Metadata \\ \hline 
botometer-feedback-2019 \cite{yang2019arming} & 529 (27) & Y & Y & P & P & P \\ \hline 
botwiki-2019 \cite{yang2019arming} & 704 (100) & Y & Y & P & P & P \\ \hline 
cresci-rtbust-2019 \cite{mazza2019rtbust} & 759 (52) & Y & Y & P & P & P\\ \hline 
cresci-stock-2018 \cite{cresci2018fake} & 25987 (71) & Y & Y & P & P & P\\ \hline 
midterms-2018 \cite{yang2020scalable} & 50538 (84) & Y & Y & P & P & P\\ \hline 
political-bots-2019 \cite{yang2019arming} & 62 (100) & Y & Y & Y & N & Y\\ \hline 
reddit-2022 & 667 (75) & Y & N & Y & Y & Y \\ \hline 
instagram-2022 & 1862 (100) & Y & Y & P & N & Y \\ \hline
\end{tabular}
\caption{Statistics of datasets used. We use the aggregation of these datasets to construct a bot detection algorithm. Many of the datasets contain partial data due to the unavailability of data at collection time. Therefore, our bot detection algorithm needs to be able to handle incomplete data and provide its best-guess prediction. \newline Y: Data field present for all users \newline N: Data field not present for all users \newline P: Data field present for partial subset of users}
\label{tab:dataset_statistics}
\end{table}

\subsection{BotBuster For Everyone: Ensemble Bot Detection}
\begin{figure}[h!]
\includegraphics[width=1.1\textwidth]{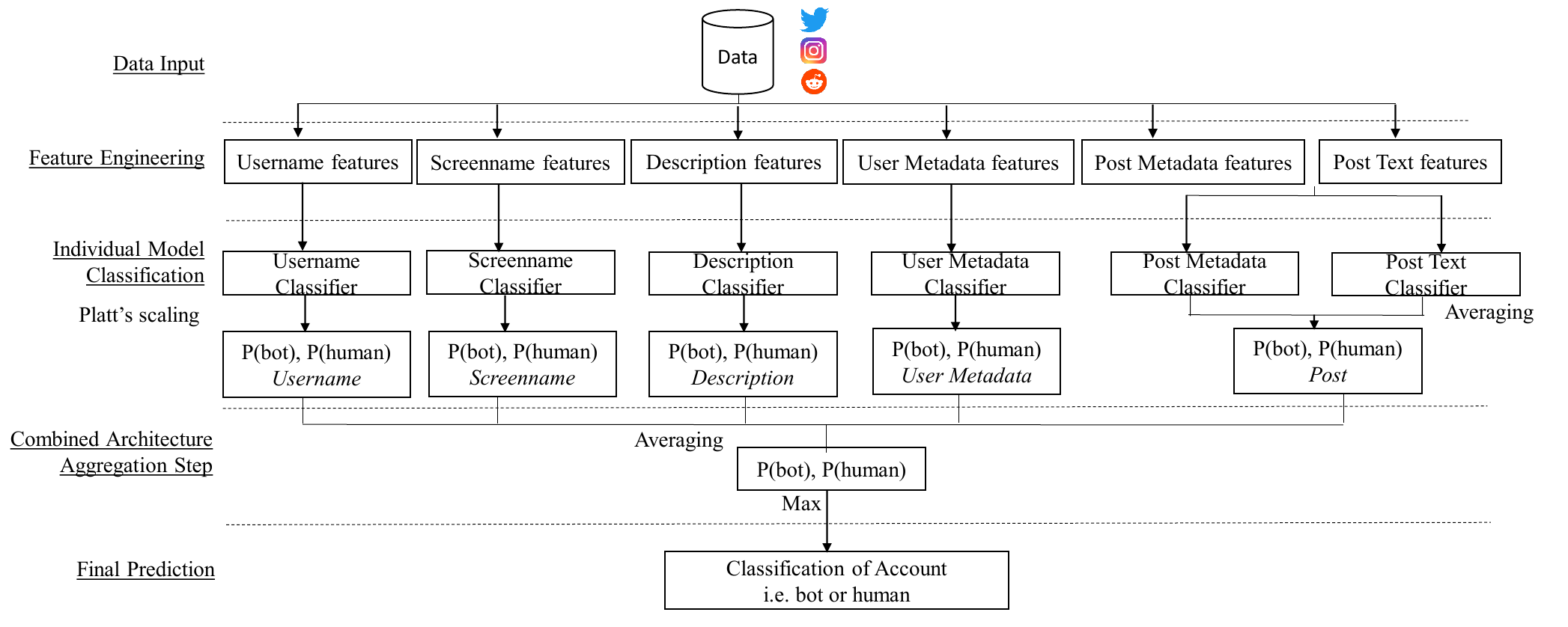}
\caption{Diagram of multi-platform bot detection ensemble. The ensemble is made up of six classifiers which extract and train/test on specialized features, providing a probability of bot/human. The probabilities are then aggregated together before the account's classification is determined by the higher of the two bot/human values.}
\label{fig:architecture}
\end{figure}

We propose an ensemble method for multi-platform bot detection. This method is illustrated in \autoref{fig:architecture}. The bot detection pipeline in our proposed BotBuster For Everyone contains of five steps: data input, feature engineering, individual model classification, combined aggregation and final prediction. Each of the steps are described in further detail in the following subsections. 

Our pipeline first begins with data input and feature engineering, to format the data from social media accounts and extract the salient features.
Model training and testing is implemented using the scikit-learn Python package. Our ensemble method involves a two-step training/testing strategy: the individual model construction and the combined architecture construction. We use the accuracy score as an evaluation metric, in order to focus on correctly classified observations of both bot and human classes. 

\paragraph{\textbf{Data Input}}
The data input step reads and processes user data, conforming the field names from each social media platforms to a common field mapping, thus dealing with multi-platform bot detection through commonality of data fields (i.e., a user on every platform has a userid and a username). It also provides an identifier at the initialization step to indicate which type of platform the data is being drawn from, so that the rest of the bot detection procedure can make use of the corresponding classifiers.

\paragraph{\textbf{Feature Engineering}}
The feature engineering step extracts a set of attribute from each data field for subsequent input into the field-specific classifiers. By tuning a classifier specific to each field, we are able to use a small set of features per field, keeping feature extraction and prediction time short. \autoref{tab:forests_metrics} includes a summary of the features extracted in this feature engineering step to be used by the individual models.

The features that are extracted from the social media accounts for use in our bot detection model are:
\begin{enumerate}
    \item \textit{Username:} A username is a singular unique word that identifies an account. It has been successfully used by its own to classify bots \cite{beskow2019its}. We distill the username into the number of uppercase and lowercase letters and the number of digits and punctuations, and the measure of string entropy. We will elaborate on the calculation of string entropy for usernames later.
    \item \textit{Screenname:} Screennames are a longer name identifier for a user, and can contain multiple words and emojis. We use the same features as the Username field, but include the number of emojis, hashtags and words as additional features. Similar to username, we distill the screenname into the number of uppercase and lowercase letters and the number of digits and punctuations, and the measure of string entropy.  We will elaborate on the calculation of string entropy for screennames later.
    \item \textit{Description:} The description is a short excerpt the user writes of himself. This field is broken down into words and the Term Frequency–Inverse Document Frequency (TFIDF) statistic of each word across the corpus of user descriptions are used as features.
    \item \textit{User Metadata:} User-based features have been successfully used in bot classification \cite{ferrara2016rise}. We use the followers count, following count, total post count, total likes count and indicators of verified and protected accounts, if available from the data. 
    \item \textit{Posts:} Data from each post is typically in the form of continuous text. This text is split into the main text and the corresponding post metadata. The main text is processed using the TFIDF statistic, while the metadata (number of likes, retweets, replies, quotes) is captured as a series of integers.
\end{enumerate}

To calculate the entropy of usernames/screennames, we first collected 3.8million names from users who posted in the last year. Using this corpus of names, we constructed a frequency of characters used in the usernames, before using the dictionary to calculate username entropy. For a username $X$, its entropy $H(X)$ is: $H(X) = -\sum^n_{i=1}{P(x_i)}log_2P(x_i)$, where $P(x_i)$ is the probability of the $i$th of the $n$ characters appearing in the username. We curated our own list of names because the probability distributions of characters in social media usernames can differ from that of the English dictionary. Screenname entropy is calculated similarly. 

\paragraph{\textbf{Individual Model Classification}}
In our first step of training/testing, we constructed individual classification models for each data field. To do so, we first combine all the datasets to form a meta-training dataset in the following fashion. For each dataset, we partition it with a 80-20 train-test split with stratification by bot/human class. This ensures that there is the same proportion of bots/humans in each training/testing set. We chose to use a supervised classification approach because our datasets have already been collected and annotated by different groups of experts, and the use of supervised classification model means that the expected output is known beforehand.

All the training splits from each dataset are then combined into a meta-training dataset used for training the individual models, and the testing splits are combined into a meta-testing dataset for testing the individual models. Then, we performed experimentation across several tree-based classifiers: decision tree, random forest, gradient boosted trees and ada boosted trees. We selected these classifiers due to their speed, which will serve advantageous in bot/human classification of large-scale datasets when deployed in actual analysis studies. Past analysis studies that characterize bot activity have used datasets that are of sizes from 40,000 \cite{uyheng2021active} to 240,000 \cite{Luceri_Deb_Giordano_Ferrara_2019} to 2.7 million users \cite{Ferrara_2017}. With such dataset sizes, speed of classification is of concern when designing a bot detection model. Five-fold cross-validation is used in all our experiments and the average results are reported.

This step takes in features specific to each data field and run them through field-specific models. Each model returns a prediction of (bot, human) tuple, which contains two values, representing the probability of the user being a bot and a human respectively. Many bot detection classifiers make use of a threshold-based classification. If the resultant probability of a bot is above a certain threshold, the social media account is classified as a bot; if the resultant probability is below the same threshold, the social media account is classified as a human. Since the choice of the threshold value can affect the percentage of bots identified, we determined the bot/human class of the user through the higher of two values that represent the probability of a bot and a human.

Separating the data input for different classifiers enables the fine-tuning of classification models specific to the feature set of each data field. This structure also deals with the problem of incomplete data for a user. In the case of incomplete data, the pipeline performs classification using the rest of the individual classifiers for the data fields present. The classifier for the missing field returns Null values. For example, if 3 data fields were present, the 3 corresponding classifiers will return a (bot, human) tuple, while 2 classifiers return null values.

Each individual model is then evaluated based on their overall accuracy and their ability to generalize across social media platforms. We selected the best model for each data field as the one with the highest accuracy and does not give a 0\% accuracy score for the Reddit and Instagram datasets. After measuring overall accuracy on the meta-testing dataset, we partitioned out the non-Twitter users and perform an evaluation on them to quantify the models' accuracy on these datasets. This method sometimes sacrifices a little overall accuracy but ensures multi-platform generalizability. For example, for classification using the Description data field, the Gradient Boosting classifier performs the best at 81.59\% accuracy, but it gives a 0\% accuracy for the Instagram dataset. We then selected the Decision Tree classifier which performed at 70.48\% overall accuracy but gave a non-zero accuracy for the Instagram dataset, and so will be able to evaluate user accounts originating from the Instagram platform. 

The final set of chosen classifiers are: decision tree for username, screenname and description, gradient boosting classifier for user metadata and random forest for posts. After choosing these best classifiers, we retrained the individual models and their outputs are calibrated using Platt's scaling, adapting the idea from past work on specialized ensembles for different types of bots \cite{sayyadiharikandeh2020detection}. This scaling is implemented using the Calibrated Classifier function\footnote{\url{https://scikit-learn.org/stable/modules/generated/sklearn.calibration.CalibratedClassifierCV.html}}. Platt's scaling calibrates the outputs of each classifier into a probability distribution using logistic regression. This therefore makes the probability returned in the (bot, human) tuple in each of the six classifiers comparable.

\paragraph{\textbf{Combined Aggregation}}
The combined aggregation step aggregates the non-null (bot, human) probability scores for the individual classifiers. The final bot classification is determined by the larger of the values in the final (bot, human) tuple. In this fashion, the need for determining a suitable threshold to classify whether an account is a bot or a human is eliminated, reducing ambiguity of the classification. This step creates the ensemble model, combining the different individual models together, and can therefore better generalize data features as a whole \cite{sayyadiharikandeh2020detection}.

No further model training is required to combine the individual models. The bot and human probabilities generated by the individual models are averaged out to produce a final bot/human probability. Testing occurs dataset by dataset, where all accounts of each dataset are evaluated for its bot probability, and the final accuracy is reported. We perform two evaluation metrics: the first, we evaluate model accuracy on the data points that the model can process, ignoring unprocessed data points. The second, we set the prediction of users that cannot be processed as the ``human" class before making an overall accuracy comparison. This mimics the use of bot detection algorithms in analysis: any user not marked as a ``bot" is typically not considered when analyzing bot behavior, and thus treated as ``humans".

Although this method of evaluation means that some data points were previously seen by the algorithms, this preserves evaluation consistency as the baseline algorithms are also trained on the same datasets and we are unable to perform separate testing on them.

\begin{table*}[!hpt]
\centering
\begin{tabular}{|p{1.8cm}|p{2.3cm}|p{1.5cm}|p{1.5cm}|p{1.5cm}|p{1.5cm}|}
\hline
\textbf{Data} & \textbf{Features Used} & \textbf{Decision Tree} & \textbf{Random Forest} & \textbf{Gradient Boosting} & \textbf{Ada Boost} \\ \hline 
Username & string entropy, \#uppercase letters, \#lowercase letters, \#digits, \#punctuations, \#emojis, \#hashtags & \textbf{75.81}$^{R,IG}$ & 75.94$^{IG}$ & 72.72$^{IG}$ & 72.14$^{R,IG}$ \\ \hline 
Screenname & string entropy, \#uppercase letters, \#lowercase letters, \#digits, \#punctuations, \#emojis, \#hashtags, \#words & \textbf{75.69}$^{R,IG}$ & 79.54$^{IG}$ & 72.08$^{R,IG}$ & 71.57$^{IG}$\\ \hline 
Description* & TF-IDF & \textbf{70.48}$^{IG}$ & 69.84 & 81.59 & 79.26 \\ \hline
User Metadata & \#followers, \#following, \#listed, \#posts, \#likes, protected, verified & 100$^{IG}$ & 74.80$^{IG}$ & \textbf{100}$^{R,IG}$ & 100$^{R,IG}$\\ \hline 
Posts** & TF-IDF, \#likes, \#retweets, \#replies, \#quotes & 56.37 & \textbf{81.02}$^{R}$ & 79.97 & 79.02$^{R}$\\
\hline
\end{tabular}
\caption{Accuracy metrics of individual models. The final ensemble combination selected are highlighted in bold. \newline *No description data available for Reddit. \newline **No posts data available for Instagram. \newline R: Model gives non-zero accuracy for Reddit dataset \newline IG: Model gives non-zero accuracy for Instagram dataset}
\label{tab:forests_metrics}
\end{table*}

\subsection{Model Evaluation}

\paragraph{\textbf{Baseline Algorithms}}
We compare our bot detection algorithm implementation against two commonly employed algorithms: BotHunter \cite{beskow2018bot} and Botometer \cite{yang2019arming}. Both algorithms are constructed using random forests for Twitter data. BotHunter uses a tiered approach that includes more user and content features as the tiers progress. Botometer is an ensemble method that relies on the real-time query of the profile. Botometer was also used by Elon Musk when estimating the proportion of Twitter bots \cite{clayton_2022}.

\paragraph{\textbf{Combined Aggregation Evaluation}}
On average, our combined framework performed with an overall accuracy of 75.47\% across the nine datasets. A summary of our results are presented in \autoref{tab:results}, which presents the overall accuracy and the percentage of the dataset each algorithm processed.  The overall accuracy is calculated by assuming that the unprocessed data are humans. This assumption mimics how one typically uses bot detection in an analysis: zooming in on the positively identified bot users and analyzing the rest of the users as a human or non-bot class separately. A more detailed summary of the results are presented in \autoref{tab:full_accuracy_botbuster}, together with other accuracy metrics that account for proportion of bot/humans in the training and testing datasets.

Our bot detection framework outperforms the overall accuracy of both baselines, which fares at 35.92\% accuracy for the BotHunter baseline and 31.46\% accuracy for Botometer baselines. The better performance of our ensemble framework can be attributed to the fact that it can process partial data and data on non-Twitter platforms. While selecting individual models, there were some cases we sacrificed accuracy for non-zero accuracy on Reddit/Instagram dataset. The accuracy of each type of individual model is reported in \autoref{tab:forests_metrics}. This shows that not all model architectures are equally adept at differentiating bot/human features across platforms, and can be overwhelmed by the larger volume of Twitter data. However, the accuracy scores of Reddit and Instagram datasets are lower, indicating that bot features may be slightly different on the three social media platforms.

Both baseline algorithms are unable to process partial data: BotHunter relies on the complete field set while Botometer relies on the survival of the account. However, our method breaks up the data into chunks for processing, allowing evaluation based on the available data. This is useful in the case of incomplete data collection or unavailable user during collection. Separating the classifiers into individual classifiers for each data field enables optimization of models for each data field, and allows us to piece together an ensemble of different types of classifiers.

\begin{table*}[!hpt]
\centering
\begin{tabular}{|p{2.5cm}|p{2.8cm}|p{2.8cm}|p{2.8cm}|}
\hline
\textbf{Dataset} & \textbf{BotHunter} & \textbf{Botometer} & \textbf{BotBuster For Everyone} \\ \hline 
& Overall accuracy \newline (\% processed) & Overall accuracy \newline (\% processed) & Overall accuracy \newline (\% processed) \\ \hline 
botometer-feedback-2019 & 57.60 (61.44) & 59.05 (71.07) & \textbf{83.08} (100) \\ \hline 
botwiki-2019 & 53.12 (90.34) & 48.12 (92.90) & \textbf{91.60}  (100) \\ \hline 
cresci-rtbust-2019 & 61.89 (74.97) & 69.43 (78.78) & \textbf{71.65} (100) \\ \hline 
cresci-stock-2018 & 37.20 (40.57) & 39.25 (47.03) & \textbf{74.61} (100) \\ \hline 
midterms-2018 & 13.20 (11.26) & 14.15 (1.31) & \textbf{85.23} (100) \\ \hline 
political-bots-2019 & 0 (0) & 17.33(20.60) & \textbf{74.54} (100) \\ \hline 
verified-2019 & 88.60 (100) & 35.50 (98.15) & \textbf{99.57} (100) \\ \hline 
reddit-2022 & 0.30 (0) & 0 (0) & \textbf{35.68} (100) \\ \hline 
instagram-2022 & 0 (0) & 0 (0) & \textbf{60.26} (100) \\ \hline 
\textbf{Average} & 34.62 (42.06) & 31.42(45.52) & \textbf{75.14} (100) \\
\hline
\end{tabular}
\caption{Summary of Results for Bot Detection Algorithms. BotHunter and Botometer are unable to process all the data, while BotBuster for Everyone method is able to. The overall accuracy is calculated by assuming the unprocessed data are humans. }
\label{tab:results}
\end{table*}

\paragraph{\textbf{External Evaluation}}
To ensure the robustness of our bot detector, we perform an evaluation on an external dataset, a dataset that has not been used in the model training before. This measures how well the model does on a dataset which it has not seen the features before, adding value to its ability to perform on out-of-domain datasets. 

We use Twibot-20 dataset \cite{feng2021twibot} for this external evaluation. the Twibot-20 was collected in 2020 via a snowball sampling method from seed users across politics, business, entertainment and sports. 

BotBuster For Everyone evaluated all the data points in the Twibot dataset and performed at 57.32\% accuracy. Our architecture outperforms the baseline BotHunter and Botometer algorithms in terms of accuracy. In terms of the number of data points processed, BotBuster For Everyone is able to process all data points unlike BotHunter and Botometer, which are not able to evaluate all the data points due to missing data fields. \autoref{tab:twibot_evaluation} presents the statistics of the evaluation ran on the Twibot-20 dataset.

\begin{table}[!ht]
\centering
\begin{tabular}{|p{3.3cm}|p{2cm}|p{2cm}|p{1.5cm}|p{1.5cm}|}
\hline
\textbf{Algorithm} & \textbf{\% processed} & \textbf{Overall Accuracy} & \textbf{MicroF1 Score} & \textbf{MacroF1 Score} \\ \hline 
BotHunter & 99.15 & 45.98 & 41.48 & 49.02 \\
Botometer & 91.38 & 33.02 & 35.98 & 30.86 \\ 
BotBuster For Everyone & 100 & 57.32 & 44.62 & 50.69 \\ \hline
\end{tabular}
\caption{Twibot-20 evaluation}
\label{tab:twibot_evaluation}
\end{table}

\paragraph{\textbf{Full Data Fields Evaluation}}
We also ask the question of how much does performance decrease in the ensemble classifier with all features in the input. For each dataset, we extracted out data points that have all the data items and ran the ensemble algorithm on those items. We report the accuracy of the ensemble classifier based on the proportion of correctly classified users out of the number of users that we are able to obtain the complete data set. We note that BotHunter and Botometer only processes data points with full data, while our method can process incomplete data.

\begin{table*}[!hpt]
\centering
\begin{tabular}{|p{4cm}|p{2.8cm}|p{2.8cm}|p{2.8cm}|}
\hline
\textbf{Dataset} & \textbf{Full Data Fields \newline (\% processed)} & \textbf{All Data Points (\% processed)} \\ \hline 
botometer-feedback-2019 & 86.55 (54.30) & 83.08 (100) \\ \hline 
botwiki-2019 & 95.20 (84.94) & 91.60 (100) \\ \hline 
cresci-rtbust-2019 & 73.67 (41.90) & 71.65 (100) \\ \hline 
cresci-stock-2018 & 78.54 (26.40) & 74.61 (100) \\ \hline 
midterms-2018 & 87.93 (14.55) & 85.23 (100) \\ \hline  
political-bots-2019 & NA (0) & 74.54 (100) \\ \hline 
verified-2019 & 99.67 (88.45) & 99.57 (100) \\ \hline 
reddit-2022 & NA (0) & 35.68 (100) \\ \hline 
instagram-2022 & NA (0) & 60.26 (100) \\ \hline 
\textbf{Average} & 86.93 (51.76) & 75.14 (100) \\
\hline
\end{tabular}
\caption{Summary of Results for BotBuster For Everyone for processing data points with full data fields. The bot detection setup does not lose a large amount of accuracy for handling incomplete data, yet it does improve the range of data that can be analyzed by the bot detector.}
\label{tab:results_full}
\end{table*}

\autoref{tab:results_full} tabulates the accuracy of the ensemble algorithm where the full data can be evaluated. Although the accuracy of the classifier with full data is higher, but because of the proportion of users with incomplete data, we think that it is worth sacrificing greater accuracy to exploit incomplete data. Even so, the individual models are trained on data points where the data for that model is available, hence it is akin to building a classifier for the full data. In addition, if we were to require all data fields to be present before performing a classification, some datasets will not be analyzed. For example, Reddit does not have a ``screen\_name", which is required for most bot detection classifiers. Therefore, this breaks the ability of our bot detection classifier to handle multiple platforms. This analysis lends weight to the architecture of our bot detector for being constructed to handle incomplete data fields: it does not lose a large amount of accuracy for handling incomplete data, yet it does improve the range of data that can be analyzed by the bot detector.

\subsection{Feature Importance Analysis}
In our feature extraction implementation, we kept the feature spaces small. Despite these, we are still able to achieve decent algorithm accuracy, showcasing that bot detection need only rely on a few key features for a decent accuracy. This provides directions for further bot account analysis: characterizing the defining features of bot accounts in contrast to human accounts. 

\begin{figure}
    \centering
    \subfigure[Username/Screenname features]{\includegraphics[width=0.70\textwidth]{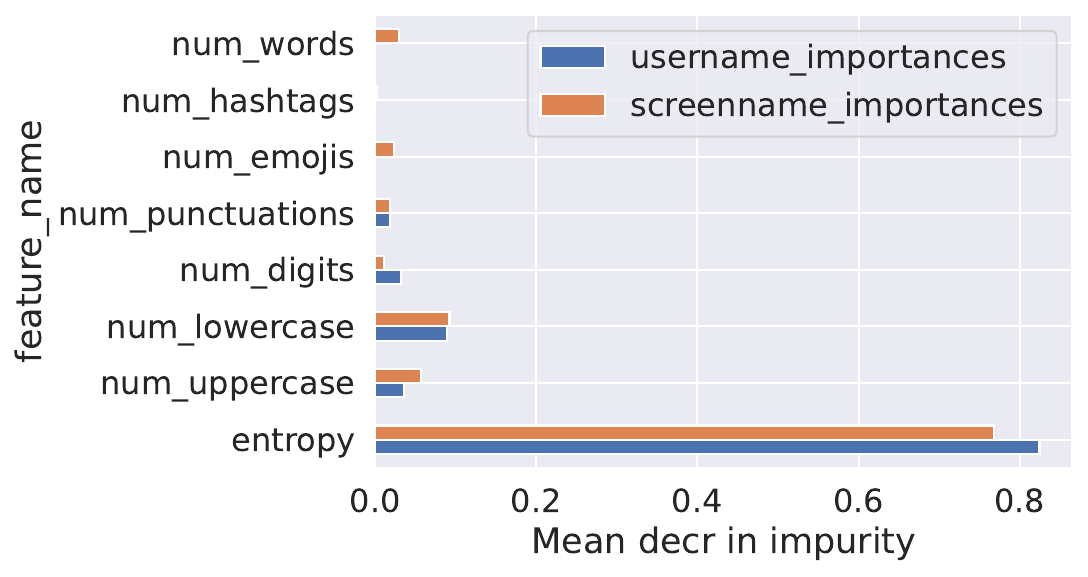}}   
    \subfigure[Posts metadata features]{\includegraphics[width=0.70\textwidth]{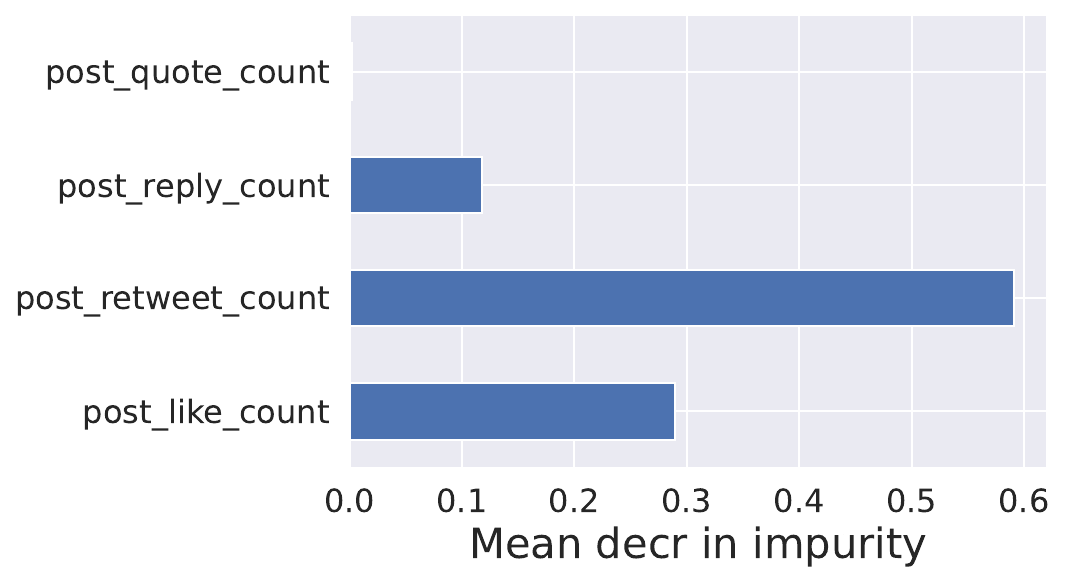}} 
    \subfigure[Description features]{\includegraphics[width=0.70\textwidth]{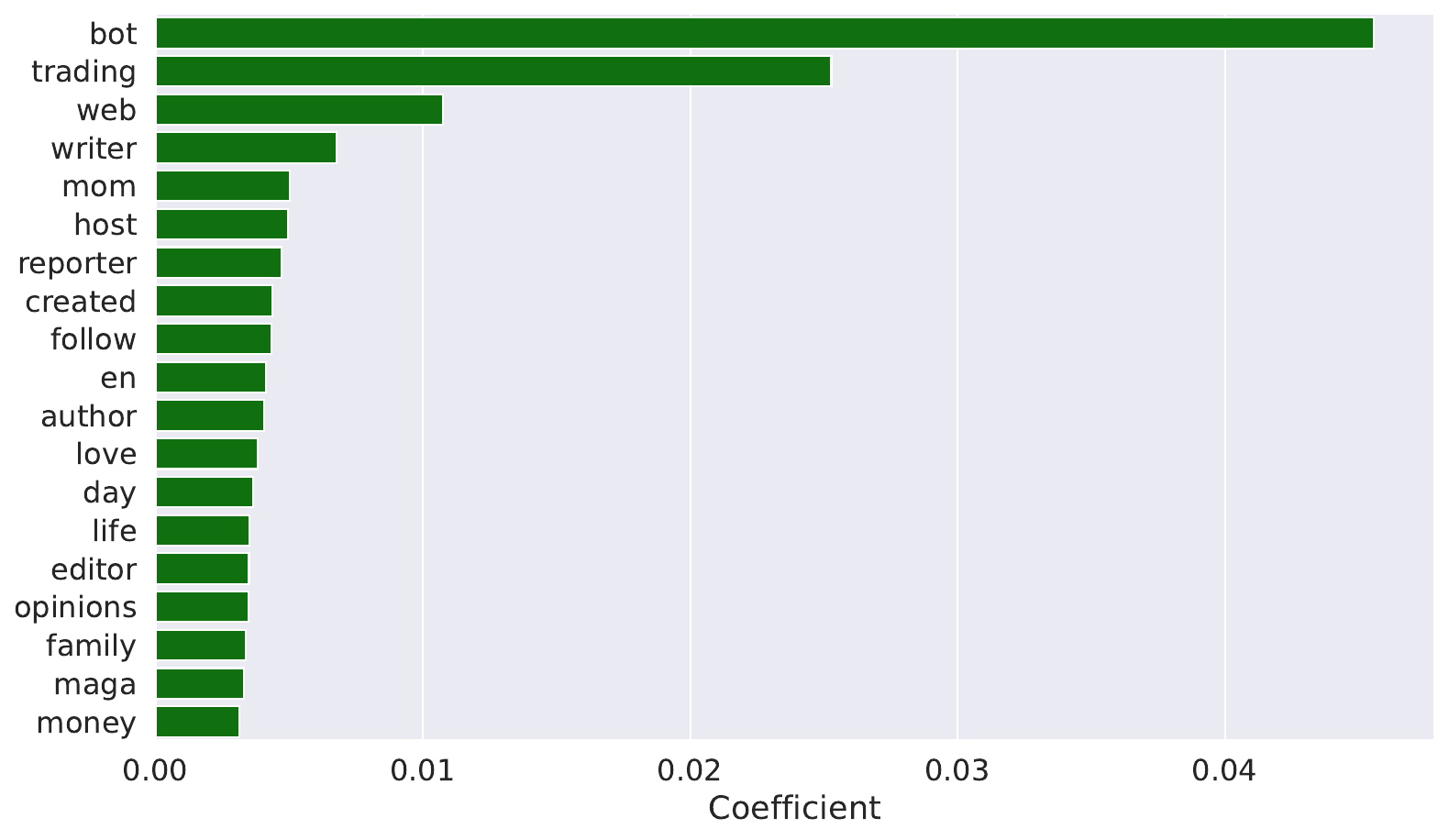}} 
    \caption{Feature Importances. The most indicative feature of bot classification is the number of retweets/shares a post receives, followed by the number of likes and the number of replies.}
    \label{fig:featureimportances}
\end{figure}

We make some observations to our feature space in the username, screenname, post metadata and description classifiers. We extracted the feature importances of each of the estimators stored in Python's sklearn classifiers of the best performing classifiers for each data class. We graphed the results in \autoref{fig:featureimportances}. 

For the username/screenname and post metadata features, they are numeric features, and hence the tree-based classifiers separate them through the decrease in impurity. The mean decrease in impurity calculates feature importances as the sum over the number of splits across the tree-based classifier. The higher mean decrease in impurity, the more important the feature is in differentiating the final bot/human class. For username/screenname, the entropy of the name string plays a large factor in the determination of bot classification. This is consistent with previous studies that characterized the randomness of profile names as an indicator of automation \cite{beskow2019its}. The R package Tweetbotornot primarily evaluates the bot likeliness of a user based on its username \cite{githubGitHubMkearneytweetbotornot}. It is also observed that the presence of digits in usernames and emojis in screennames are indications of bot/human classes of the account.

In terms of post metadata (\autoref{fig:featureimportances}), we observe that the most indicative feature of a bot classification is the number of retweets/shares a post receives, followed by the number of likes and the number of replies. This is consistent with the feature analysis of bot detection algorithm MulBot where retweets and replies are the more important features \cite{mannocci2022mulbot}. This means that posts by bot accounts have a lot more shares than human accounts, possibly pointing to their ability to construct more viral posts or indications of bot networks working together to increase influences of posts of other bots within the network.

The description of an author is a string of words, and hence is treated differently by the Decision Tree classifier. In constructing the classifier, the description string is broken down into a bag of words, therefore the feature importances of the words are represented by coefficients, where the coefficient scores how important the word is within a description string. The first word is ``bot", suggesting the incorporation of a heuristic to identify key signals of bot accounts such as words present in the description or account name \cite{Livingstone_2022}. Words representing a person's identity (i.e. writer, mom, host, author, reporter, editor etc.) are extremely indicative words, suggesting connections between the expression of identities and bot likeliness of an account. This opens avenues for further investigation on the correlation on identity expression and automation.  

\section{Application of Social Bot Detector}
\label{sec:application}
With the construction of the ensemble bot detection algorithm which we named BotBuster For Everyone, we applied it to a slice of the online discourse extracted from the US 2020 Elections from Twitter and Reddit. The 2020 United States presidential elections was held on 3 November 2020. In this election, Democratic president Joe Biden defeated incumbent Republican president Donald Trump. After the win by Biden, Trump and his supporters did not concede, and claimed voter and election fraud. Past research have been done to analyze users that have similarities across both platforms surround this incident of protest against voter fraud, and having bot detection capabilities can enhance these analysis by providing the perspective of the degree of automation of these users \cite{murdock2023identifying}.

We perform a short study on the discourse of the protest of election voter fraud. We collected social media conversations for a week after the elections, from 3 November to 9 November 2020, analyzing the bot activity within this timeframe. 

The Twitter data was collected with the Twitter V1 API and the Reddit data using the Pushshift API \cite{baumgartner2020pushshift}. For Twitter data, we have 4,351,111 unique posts and 1,183,313 unique users. The discourse was not as active on Reddit, and we retrieved a smaller amount of data from Reddit, collecting a total of 4403 unique posts and 2449 unique users.

\begin{figure}[htbp]
\centering
\includegraphics[scale=0.7]{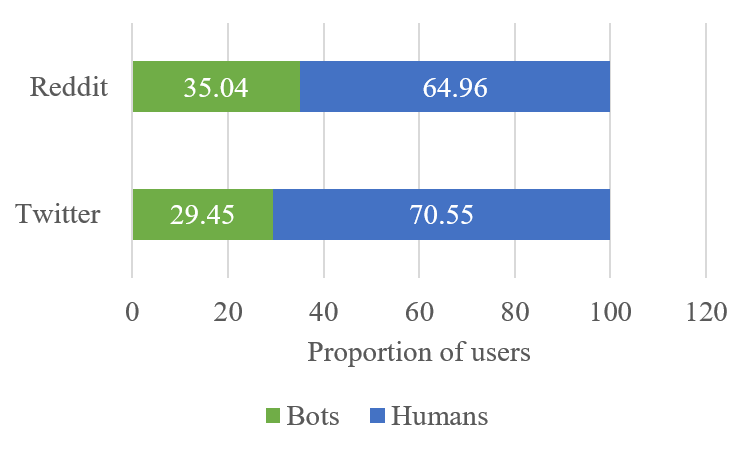}
\caption{Proportion of user types present in the US 2020 presidential elections. There is a higher proportion of bot users in Reddit than in Twitter.}
\label{fig:proportionofusers}
\end{figure}

We apply our constructed BotBuster for Everyone to identify bot users in the datasets. The same bot detection model is applied to both Twitter and Reddit data, extracting and analyzing the bots present in both datasets. \autoref{fig:proportionofusers} shows the proportion of user types extracted from these datasets. There is a higher proportion of bot users that are present in the Reddit conversation (35.04\%) as compared to Twitter (29.45\%). This observation shows that the election discussion by bots is focused on Reddit, in which the subreddit and reply structures do facilitate discussions, as compared to Twitter discussions.

We then separated the post texts written by each type of user in each of the platforms, and perform Latent Dirichlet Allocation (LDA) to identify the key narrative themes within the texts. The LDA algorithm returns a list of keywords relating to each theme, after which, the authors manually looked through and interpreted the themes, combining them where necessary. Six key themes emerged among the discourse: voter fraud in specific states, in particular the hotly contested states of Michigan, Neveda and Georgia; voter fraud in general; voter fraud from mail-in ballots; news regarding the elections and protest; the questioning of the integrity of the elections; and the call to fighting voter fraud, in particular using the catchphrase ``stop the steal".

\begin{figure}[htbp]
\centering
\includegraphics[scale=0.7]{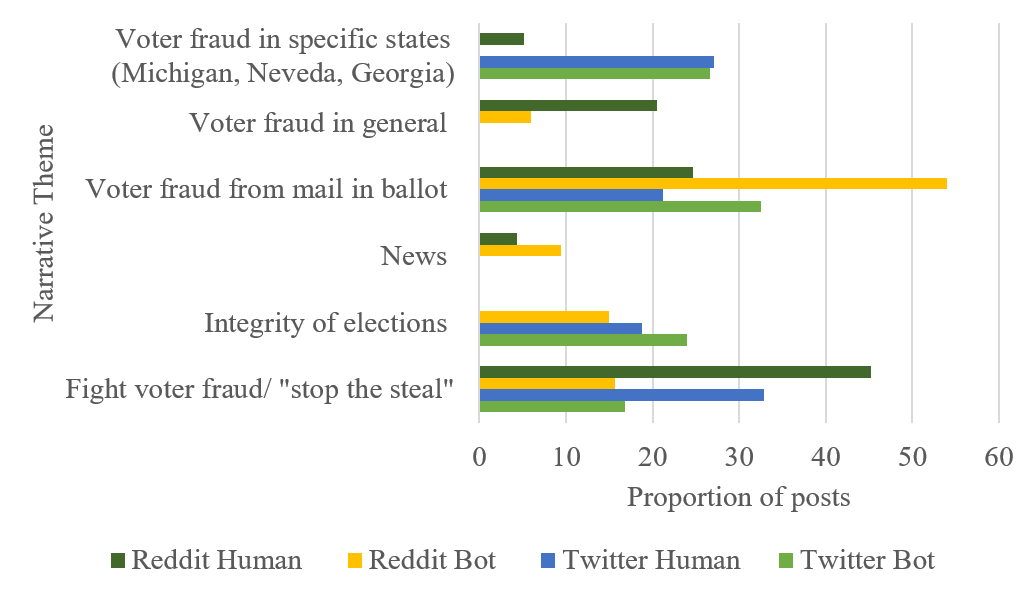}
\caption{Proportion of Narrative Themes present per user type in the US 2020 presidential elections. There are different focuses of each of the user class: bots disseminate information, while human users advocate for action.}
\label{fig:narrativethemes}
\end{figure}

\autoref{fig:narrativethemes} represents the proportion of posts by narrative themes. The narrative of voter fraud from mail in ballot and the call to fight voter fraud is present throughout both platforms, and echoed by both classes of users. The narrative of voter fraud by mail-in ballot is most echoed by Reddit bots followed by Twitter bots, while the calling out to fight voter fraud is most echoed by Reddit humans then Twitter humans. This shows the different focus of each of the user class: bots disseminate disinformation, i.e. insinuating that mailed-in ballots were rigged, and thus the elections were rigged; while human users advocate for action.

In this work, we analyzed only a small proportion of online discourse using our multi-platform bot detection model as an illustration that the model can be used to identify bots on multiple social media platforms. A subsequent step stemming from this identification analysis is to include a study into the social media accounts identified as bots and their differences with human accounts, however, that is outside the scope of this paper.

\section{Discussion}
\label{sec:discussion}
In this work, we built BotBuster For Everyone, a multi-platform social media bot detector. This model identifies bot accounts from three main platforms: Twitter, Reddit and Instagram. The input format for these platforms that are currently built-in are: Twitter V1 API, Twitter V2 API, Reddit Pushshift API and Instagram data from CrowdTangle. There is also a ``custom" format option, where users can edit a JSON file to specify the mapping between the field names of their data to the bot detector's input fields.

\paragraph{\textbf{Handling Incomplete Data Fields}}
One highlight of our bot detection model is its ability to process data where not all fields are present. In our experiment of Full Data Fields Evaluation, we observe that while the bot detection algorithm performs better when processing data where the full data fields are present, it only does so slightly better. Therefore, we put forth that it is worth sacrificing a little accuracy for a wider use of the bot detection algorithm. Further, the ability to handle incomplete data fields lends the algorithm the ability to handle multiple platforms, for our experiments show that Twitter has the larger feature set that can be extracted from the platform, while other platforms have a smaller feature set. The ability to process datasets where the data features are missing for some data points or are not present as fields in the platform allows analysis of a lot more data points especially historically collected data, and reduces the collection burden.

\paragraph{\textbf{Multi-Platform Generalizability}}
Our bot detection model can identify bots from multiple social media platforms, reducing the need to source and run multiple bot detection models for cross-platform studies, thus saving time and aggregating results. Leveraging on the fact that bots across different social media platforms have similar features, we are able to generalize our framework across three social media platforms. While there have been many bot detectors built for the Twitter platform, there are very few built for Reddit and Instagram. Our model contributes to the small set of bot detectors built for Reddit and Instagram, while aggregating training data from the bot detection repositories for Twitter. The aggregation step combines heterogenous datasets as training data which teaches the models bot/human features at different time periods and behavioral patterns, making the model more generalizable to detect different types of bots \cite{hayawi2022deeprobot}. With the ability to analyze multiple platforms, our bot detection model thus provides the opportunity to perform cross-platform user and discourse analysis on social media.

Bot accounts are also observed to work together as sophisticated and coordinated bot networks, which have been observed in the 2021 French protests \cite{10.3389/fdata.2023.1221744}, and during the 2014 Crimean water crisis \cite{khaund2021social}. Literature on identification of bot networks usually involve identifying bots singly before inferring their coordination with each other \cite{khaund2021social,pacheco2021uncovering}. Because of the coordination features, where the group of bots are motivated by a single intent, they leave behind more automation than single bots, which allows detection through bot detection mechanisms \cite{cresci2020decade}. As such, our ensemble algorithm can aid in the identification of bot networks through the initial step of bot classification of single users.

\paragraph{\textbf{Specialized fine-tuned classifiers}}
Our ensemble-based bot detection framework that fine-tunes specialized classifiers for each data class before aggregating the probabilities. This means that each classifier is specially designed to fit for the corresponding data class, making it more accurate for the data input from the data class.

The separate fine-tuning mechanism allows the overall bot detection architecture to handle cases of incomplete data. When there is incomplete data, that is, the data fields are not present in the input data, the corresponding specialized classifiers are unable to make a prediction. The rest of the specialized classifiers can still make a prediction on the prevailing data as they are trained separately and are analyzing different data fields, thus being unaffected by the lack of data from one field. This provides the ability of our bot detection model to provide predictions for as much users as possible, rather than only users with the complete suite of data fields.

Separating the classifiers also allows further interpretability of each classifier. Through analyzing each of the classifiers, we can highlight indicative features of bot accounts, such as randomness of usernames and the presence of identity terms within a user's description. 

In addition, the use of tree-based algorithms in the construction of the ensembles illustrate the principle of Occam's razor: that the models can be simple enough to be valid. Our aggregation of tree-based classifiers perform almost as well as deep-learning based classifiers, e.g., an accuracy score of 71.65\% for the cresci-rtbust-2019 dataset vs 72\% by LSTM-based model DeeProBot \cite{hayawi2022deeprobot}; 83.08\% for the botometer-feedback-2019 dataset vs 78.4\% for a concatenation of multiple LSTM models \cite{arin2023deep}. With similar accuracy ranges, we infer that in terms of differentiating bot and human classes and features, simpler classifiers work equally as well as complex classifiers. In fact, the simplicity of tree-based classifiers means that such a bot detection algorithm can be easily run and do not require heavy GPU-processing that deep learning based ones do.

\paragraph{\textbf{Eliminating the Need for Threshold Selection}}
Reflecting both the probability of bot and human as an output of the framework not only eliminates the need for selecting a threshold value for bot/human classification. Past work has shown that the proportion of bots can differ greatly between commonly used classification thresholds (50\%, 75\%, 80\%), as much as a 15\% difference. The classification threshold means that above which the user is deemed as a bot and below which it is deemed as a human \cite{adel2022security}. The elimination of threshold value thus removes the ambiguity of the selection of users as bots, and be helpful in increasing the consistency of bot detection. This also allows an analyst to objectively see the range of bot likeliness for the account as compared to its human likeliness. While typically we use the class that is indicated as the higher of the two values for the final classification, providing both likelihood values gives the analyst a chance to decide the bot/human difference is too small and thus use other features, including manual inspection, to determine the classification. 

\paragraph{\textbf{Providing Multi-Platform Perspectives}}
We applied our bot detection model towards the US 2020 presidential elections and used it to understand the differences in discourse that happened on multiple social media platforms. This is especially useful because social media discussions are not isolated to a single platform, and to gain a full perspective of the online chatter, one must analyze multiple platforms. Our bot detector provides the capability to analyze multiple platforms at once, segregating the automated bot agents, allowing for subsequent analysis for better understanding the discussion on the event.

\paragraph{\textbf{Limitations and Future Work}}
One key limitation of constructing bot algorithms is obtaining a representative set of annotated data. Several of the Twitter datasets are skewed towards financial and election bots. The classifiers trained on this data will be overfitted towards these topics, leaving bots that operate under other themes undetected. The Reddit dataset relies on crowd-sourced ranking of bots via majority voting, which is subject to social influence. Lastly, although the purchased Instagram dataset has no ambiguity of bot classification, it contains only users of the positive bot class and requires further curation for a balanced dataset. It also only contains one type of bot, the follower bot, and more work is required to characterize and consolidate a dataset spanning different bot types.

Future work involves sampling representative bot datasets both within and across platforms to improve generalizability of the classifiers across the scope of social media. Bot/human account features on social media are continually evolving. For example, past work has observed that across the years of bot detection data development and collection, the linguistic features of posts of bots/humans have been observed to evolve \cite{ng2023botbuster}. In addition, new technologies such as generative language technologies can change the behavior and the feature set of both bot and human users \cite{arin2023deep}. Therefore, future versions of this supervised bot detector involves continual updating and training of the model in order to keep up to date with the latest bot/human account similarities and differences. We acknowledge that the limitation to supervised learning bot detection models means that researchers need to continually observe the space, but it also opens avenues for research and observation of bot/human behavior changes.

%Doing so entails identifying the similarities and differences between data features of each platform, and later tuning individual classifiers for them.

\section{Conclusion}
In this work, we constructed a multi-platform social bot detector, Botbuster For Everyone, which works through assembling an ensemble of tree-based models. Each tree model is specific to a feature extracted from a social media profile, and is individually trained. The outputs of each model are aggregated together to return a probability of whether an account is a bot or a human. Breaking down the account evaluation by features provides BotBuster for Everyone the capability to deal with incomplete data where prediction can be made using available data and classifiers, and incorporate data from multiple platforms by using similarities in data field names. The use of tree-based classification models provides the ability to perform bot detection and classification quickly on lighter-weight computing hardware with CPUs, thus increasing the access of bot detection models.

We also applied BotBuster for Everyone on a dataset of US 2020 Elections discourse, analyzing the topic differences between bot and humans on Reddit and Twitter. We found that there is a higher proportion of bot users in the collected Reddit conversation than Twitter users, and that bot accounts are typically used for (dis)information dissemination and human users for advocating action.

Bot accounts can threaten the health of our social, and even economic, systems. As such, there is a need for continual efforts in bot detection research to detect these accounts at scale. As social media discourse diversify across platforms, there must be a bot detection model that is able to seamlessly identify automated accounts across multiple platforms quickly and at scale. We designed Botbuster For Everyone, a bot detection framework which generalizes across three social media platforms. We hope that our work provides inspiration for future research on bot detection and investigation.

\backmatter

\bmhead{Supplementary information}
Supplementary files containing the full details of accuracy metrics are attached.

\bmhead{Acknowledgments}
The research for this paper was supported by the following grants: Cognizant Center of Excellence Content Moderation Research Program, Office of Naval Research (Bothunter, N000141812108), Scalable Technologies for Social Cybersecurity/ARMY (W911NF20D0002), Air Force Research Laboratory/CyberFit (FA86502126244). The views and conclusions  are those of the authors and should not be interpreted as representing the official policies, either expressed or implied.

\section*{Declarations}

\textbf{Conflict of interest/Competing interests.} The authors declare that there are no conflicting interests. 

\textbf{Availability of data and materials.} Data used for training the bot detection models are publicly available at \url{https://botometer.osome.iu.edu/bot-repository/}. Data collected for the US 2020 elections can be made available through the corresponding author, in accordance to Twitter's terms and conditions. Code to the bot detection model is available at \url{https://github.com/quarbby/BotBuster-4-Everyone}.

\textbf{Author Contributions.} L.H.X. Ng conceptualized and conducted the experiments, and wrote the manuscript. K.M. Carley reviewed the manuscript. All authors approved of the final manuscript.  

\begin{appendices}

\section{Full Accuracy Metrics}
\autoref{tab:full_accuracy_botbuster}, \autoref{tab:full_accuracy_bothunter} and  \autoref{tab:full_accuracy_botometer} presents the full set of accuracy metrics for the aggregated bot detection classifier and the baseline classifiers. 

\begin{table*}[!hpt]
\centering
\begin{tabular}{|p{3.3cm}|p{1.5cm}|p{1.5cm}|p{1.5cm}|p{1.5cm}|}
\hline
\textbf{Dataset} & \multicolumn{4}{|c|}{\textbf{BotBuster For Everyone}} \\ \hline
& \% processed & Accuracy & Micro-F1 score & Macro-F1 score\\ \hline
botometer-feedback-2019 & 100 & 83.08 & 83.08 & 78.81 \\ \hline 
botwiki-2019 & 100 & 91.60 & 34.34 & 48.55 \\ \hline 
cresci-rtbust-2019 & 100 & 71.65 & 70.83 & 70.13 \\ \hline
cresci-stock-2018 & 100 & 74.61 & 74.43 & 74.38\\ \hline
midterms-2018 & 100 & 85.23 & 89.73 & 49.46 \\ \hline
political-bots-2019 & 100 & 74.54 & 78.64 & 68.45 \\ \hline 
verified-2019 & 100 & 99.57 & 99.42 & 50.29 \\ \hline 
reddit-2022 & 100 & 34.68 & 51.20  & 33.86 \\ \hline 
instagram-2022 & 100 & 60.26 & 69.67 & 37.61\\ 
\hline
\textbf{Average} & 100 & 50.69 & 44.62 & 56.84\\ 
\hline
\end{tabular}
\caption{Summary of final results for BotBuster For Everyone algorithm. The overall accuracy is calculated by assuming the unprocessed data are humans. }
\label{tab:full_accuracy_botbuster}
\end{table*}

\begin{table*}[!hpt]
\centering
\begin{tabular}{|p{3.3cm}|p{1.5cm}|p{1.5cm}|p{1.5cm}|p{1.5cm}|p{1.5cm}|}
\hline
\textbf{Dataset} & \multicolumn{5}{|c|}{\textbf{BotHunter}} \\ \hline
& \% processed & Accuracy (processed) & Accuracy (overall) & Micro-F1 score (Overall) & Macro-F1 score (Overall) \\ \hline
botometer-feedback-2019 & 61.44 & 74.10 & 57.60 & 70.65 & 69.75\\ \hline 
botwiki-2019 & 90.34 & 53.13 & 53.12 & 69.39 & 34.69  \\ \hline 
cresci-rtbust-2019 & 74.97 & 62.90 &  91.89 & 62.98 & 65.87  \\ \hline
cresci-stock-2018 & 40.57 & 37.36 & 37.20 & 48.09 & 35.93\\ \hline
midterms-2018 & 100 & 11.26 & 15.30 & 13.20 & 9.02 \\ \hline
political-bots-2019 & 0 & 0 & 0 & 0 & 0 \\ \hline 
verified-2019 & 88.60 & 100 & 100 & 100 & 100\\ \hline 
reddit-2022 & 0 & 0 & 0 & 0 & 0 \\ \hline 
instagram-2022 & 0 & 0 & 0 & 0 & 0  \\ 
\hline
\textbf{Average} & 52.45 & 57.13 & 40.37 & 40.01 & 34.55 \\ 
\hline
\end{tabular}
\caption{Summary of final results for BotHunter algorithm. The overall accuracy is calculated by assuming the unprocessed data are humans. }
\label{tab:full_accuracy_bothunter}
\end{table*}

\begin{table*}[!hpt]
\centering
\begin{tabular}{|p{3.3cm}|p{1.5cm}|p{1.5cm}|p{1.5cm}|p{1.5cm}|p{1.5cm}|}
\hline
\textbf{Dataset} & \multicolumn{5}{|c|}{\textbf{Botometer}} \\ \hline
& \% processed & Accuracy (processed) & Accuracy (overall) & Micro-F1 score (Overall) & Macro-F1 score (Overall) \\ \hline
botometer-feedback-2019 & 71.07 & 53.68 & 59.05 & 60.95 & 63.89 \\ \hline 
botwiki-2019 & 92.90 & 92.89 & 48.12 & 50.07 & 46.37 \\ \hline 
cresci-rtbust-2019 & 78.78 & 38.12 & 69.43 & 69.42 & 69.38 \\ \hline
cresci-stock-2018 & 47.03 & 38.12 & 39.25 & 44.10 & 41.39 \\ \hline
midterms-2018 & 47.03 & 11.90 & 14.15 & 12.40 & 7.22 \\ \hline
political-bots-2019 & 1.31 & 20.60 & 17.33 & 21.48 & 20.96 \\ \hline 
verified-2019 & 20.60 & 30.20 & 35.50 & 25.71 & 27.53 \\ \hline 
reddit-2022 & 0 & 0 & 0 & 0 & 0\\ \hline 
instagram-2022 & 0 & 0 & 0 & 0 & 0 \\ 
\hline
\textbf{Average} & 45.52 & 31.72 & 31.43 & 31.57 & 30.75\\ 
\hline
\end{tabular}
\caption{Summary of final results for Botometer algorithm. The overall accuracy is calculated by assuming the unprocessed data are humans. }
\label{tab:full_accuracy_botometer}
\end{table*}

\end{appendices}

\bibliography{main}% common bib file


\begin{thebibliography}{47}
\ifx \bisbn   \undefined \def \bisbn  #1{ISBN #1}\fi
\ifx \binits  \undefined \def \binits#1{#1}\fi
\ifx \bauthor  \undefined \def \bauthor#1{#1}\fi
\ifx \batitle  \undefined \def \batitle#1{#1}\fi
\ifx \bjtitle  \undefined \def \bjtitle#1{#1}\fi
\ifx \bvolume  \undefined \def \bvolume#1{\textbf{#1}}\fi
\ifx \byear  \undefined \def \byear#1{#1}\fi
\ifx \bissue  \undefined \def \bissue#1{#1}\fi
\ifx \bfpage  \undefined \def \bfpage#1{#1}\fi
\ifx \blpage  \undefined \def \blpage #1{#1}\fi
\ifx \burl  \undefined \def \burl#1{\textsf{#1}}\fi
\ifx \doiurl  \undefined \def \doiurl#1{\url{https://doi.org/#1}}\fi
\ifx \betal  \undefined \def \betal{\textit{et al.}}\fi
\ifx \binstitute  \undefined \def \binstitute#1{#1}\fi
\ifx \binstitutionaled  \undefined \def \binstitutionaled#1{#1}\fi
\ifx \bctitle  \undefined \def \bctitle#1{#1}\fi
\ifx \beditor  \undefined \def \beditor#1{#1}\fi
\ifx \bpublisher  \undefined \def \bpublisher#1{#1}\fi
\ifx \bbtitle  \undefined \def \bbtitle#1{#1}\fi
\ifx \bedition  \undefined \def \bedition#1{#1}\fi
\ifx \bseriesno  \undefined \def \bseriesno#1{#1}\fi
\ifx \blocation  \undefined \def \blocation#1{#1}\fi
\ifx \bsertitle  \undefined \def \bsertitle#1{#1}\fi
\ifx \bsnm \undefined \def \bsnm#1{#1}\fi
\ifx \bsuffix \undefined \def \bsuffix#1{#1}\fi
\ifx \bparticle \undefined \def \bparticle#1{#1}\fi
\ifx \barticle \undefined \def \barticle#1{#1}\fi
\bibcommenthead
\ifx \bconfdate \undefined \def \bconfdate #1{#1}\fi
\ifx \botherref \undefined \def \botherref #1{#1}\fi
\ifx \url \undefined \def \url#1{\textsf{#1}}\fi
\ifx \bchapter \undefined \def \bchapter#1{#1}\fi
\ifx \bbook \undefined \def \bbook#1{#1}\fi
\ifx \bcomment \undefined \def \bcomment#1{#1}\fi
\ifx \oauthor \undefined \def \oauthor#1{#1}\fi
\ifx \citeauthoryear \undefined \def \citeauthoryear#1{#1}\fi
\ifx \endbibitem  \undefined \def \endbibitem {}\fi
\ifx \bconflocation  \undefined \def \bconflocation#1{#1}\fi
\ifx \arxivurl  \undefined \def \arxivurl#1{\textsf{#1}}\fi
\csname PreBibitemsHook\endcsname

%%% 1
\bibitem{ferrara2016rise}
\begin{barticle}
\bauthor{\bsnm{Ferrara}, \binits{E.}},
\bauthor{\bsnm{Varol}, \binits{O.}},
\bauthor{\bsnm{Davis}, \binits{C.}},
\bauthor{\bsnm{Menczer}, \binits{F.}},
\bauthor{\bsnm{Flammini}, \binits{A.}}:
\batitle{The rise of social bots}.
\bjtitle{Communications of the ACM}
\bvolume{59}(\bissue{7}),
\bfpage{96}--\blpage{104}
(\byear{2016})
\end{barticle}
\endbibitem

%%% 2
\bibitem{ng2022pro}
\begin{botherref}
\oauthor{\bsnm{Ng}, \binits{L.H.X.}},
\oauthor{\bsnm{Carley}, \binits{K.M.}}:
Pro or anti? a social influence model of online stance flipping.
IEEE Transactions on Network Science and Engineering
(2022)
\end{botherref}
\endbibitem

%%% 3
\bibitem{ferrara2016predicting}
\begin{bchapter}
\bauthor{\bsnm{Ferrara}, \binits{E.}},
\bauthor{\bsnm{Wang}, \binits{W.-Q.}},
\bauthor{\bsnm{Varol}, \binits{O.}},
\bauthor{\bsnm{Flammini}, \binits{A.}},
\bauthor{\bsnm{Galstyan}, \binits{A.}}:
\bctitle{Predicting online extremism, content adopters, and interaction
  reciprocity}.
In: \bbtitle{International Conference on Social Informatics},
pp. \bfpage{22}--\blpage{39}
(\byear{2016}).
\bcomment{Springer}
\end{bchapter}
\endbibitem

%%% 4
\bibitem{yang2019arming}
\begin{barticle}
\bauthor{\bsnm{Yang}, \binits{K.-C.}},
\bauthor{\bsnm{Varol}, \binits{O.}},
\bauthor{\bsnm{Davis}, \binits{C.A.}},
\bauthor{\bsnm{Ferrara}, \binits{E.}},
\bauthor{\bsnm{Flammini}, \binits{A.}},
\bauthor{\bsnm{Menczer}, \binits{F.}}:
\batitle{Arming the public with artificial intelligence to counter social
  bots}.
\bjtitle{Human Behavior and Emerging Technologies}
\bvolume{1}(\bissue{1}),
\bfpage{48}--\blpage{61}
(\byear{2019})
\end{barticle}
\endbibitem

%%% 5
\bibitem{rauchfleisch2020false}
\begin{barticle}
\bauthor{\bsnm{Rauchfleisch}, \binits{A.}},
\bauthor{\bsnm{Kaiser}, \binits{J.}}:
\batitle{The false positive problem of automatic bot detection in social
  science research}.
\bjtitle{PloS one}
\bvolume{15}(\bissue{10}),
\bfpage{0241045}
(\byear{2020})
\end{barticle}
\endbibitem

%%% 6
\bibitem{ng2022stabilizing}
\begin{barticle}
\bauthor{\bsnm{Ng}, \binits{L.H.X.}},
\bauthor{\bsnm{Robertson}, \binits{D.C.}},
\bauthor{\bsnm{Carley}, \binits{K.M.}}:
\batitle{Stabilizing a supervised bot detection algorithm: How much data is
  needed for consistent predictions?}
\bjtitle{Online Social Networks and Media}
\bvolume{28},
\bfpage{100198}
(\byear{2022})
\end{barticle}
\endbibitem

%%% 7
\bibitem{clayton_2022}
\begin{botherref}
\oauthor{\bsnm{Clayton}, \binits{J.}}:
Doubts cast over Elon Musk's Twitter bot claims.
BBC
(2022).
\url{https://www.bbc.com/news/technology-62571733}
\end{botherref}
\endbibitem

%%% 8
\bibitem{ng2023deflating}
\begin{barticle}
\bauthor{\bsnm{Ng}, \binits{L.H.X.}},
\bauthor{\bsnm{Carley}, \binits{K.M.}}:
\batitle{Deflating the chinese balloon: types of twitter bots in us-china
  balloon incident}.
\bjtitle{EPJ Data Science}
\bvolume{12}(\bissue{1}),
\bfpage{63}
(\byear{2023})
\end{barticle}
\endbibitem

%%% 9
\bibitem{jacobs2023tracking}
\begin{bchapter}
\bauthor{\bsnm{Jacobs}, \binits{C.S.}},
\bauthor{\bsnm{Ng}, \binits{L.H.X.}},
\bauthor{\bsnm{Carley}, \binits{K.M.}}:
\bctitle{Tracking china’s cross-strait bot networks against taiwan}.
In: \bbtitle{International Conference on Social Computing, Behavioral-cultural
  Modeling and Prediction and Behavior Representation in Modeling and
  Simulation},
pp. \bfpage{115}--\blpage{125}
(\byear{2023}).
\bcomment{Springer}
\end{bchapter}
\endbibitem

%%% 10
\bibitem{gera2022t}
\begin{barticle}
\bauthor{\bsnm{Gera}, \binits{S.}},
\bauthor{\bsnm{Sinha}, \binits{A.}}:
\batitle{T-bot: Ai-based social media bot detection model for trend-centric
  twitter network}.
\bjtitle{Social Network Analysis and Mining}
\bvolume{12}(\bissue{1}),
\bfpage{76}
(\byear{2022})
\end{barticle}
\endbibitem

%%% 11
\bibitem{yang2020scalable}
\begin{bchapter}
\bauthor{\bsnm{Yang}, \binits{K.-C.}},
\bauthor{\bsnm{Varol}, \binits{O.}},
\bauthor{\bsnm{Hui}, \binits{P.-M.}},
\bauthor{\bsnm{Menczer}, \binits{F.}}:
\bctitle{Scalable and generalizable social bot detection through data
  selection}.
In: \bbtitle{Proceedings of the AAAI Conference on Artificial Intelligence},
vol. \bseriesno{34},
pp. \bfpage{1096}--\blpage{1103}
(\byear{2020})
\end{bchapter}
\endbibitem

%%% 12
\bibitem{heidari2021empirical}
\begin{bchapter}
\bauthor{\bsnm{Heidari}, \binits{M.}},
\bauthor{\bsnm{James~Jr}, \binits{H.}},
\bauthor{\bsnm{Uzuner}, \binits{O.}}:
\bctitle{An empirical study of machine learning algorithms for social media bot
  detection}.
In: \bbtitle{2021 IEEE International IOT, Electronics and Mechatronics
  Conference (IEMTRONICS)},
pp. \bfpage{1}--\blpage{5}
(\byear{2021}).
\bcomment{IEEE}
\end{bchapter}
\endbibitem

%%% 13
\bibitem{kantepe2017preprocessing}
\begin{bchapter}
\bauthor{\bsnm{Kantepe}, \binits{M.}},
\bauthor{\bsnm{Ganiz}, \binits{M.C.}}:
\bctitle{Preprocessing framework for twitter bot detection}.
In: \bbtitle{2017 International Conference on Computer Science and Engineering
  (ubmk)},
pp. \bfpage{630}--\blpage{634}
(\byear{2017}).
\bcomment{IEEE}
\end{bchapter}
\endbibitem

%%% 14
\bibitem{pratama2019social}
\begin{barticle}
\bauthor{\bsnm{Pratama}, \binits{P.G.}},
\bauthor{\bsnm{Rakhmawati}, \binits{N.A.}}:
\batitle{Social bot detection on 2019 indonesia president candidate’s
  supporter’s tweets}.
\bjtitle{Procedia Computer Science}
\bvolume{161},
\bfpage{813}--\blpage{820}
(\byear{2019})
\end{barticle}
\endbibitem

%%% 15
\bibitem{kudugunta2018deep}
\begin{barticle}
\bauthor{\bsnm{Kudugunta}, \binits{S.}},
\bauthor{\bsnm{Ferrara}, \binits{E.}}:
\batitle{Deep neural networks for bot detection}.
\bjtitle{Information Sciences}
\bvolume{467},
\bfpage{312}--\blpage{322}
(\byear{2018})
\end{barticle}
\endbibitem

%%% 16
\bibitem{cresci2018fake}
\begin{bchapter}
\bauthor{\bsnm{Cresci}, \binits{S.}},
\bauthor{\bsnm{Lillo}, \binits{F.}},
\bauthor{\bsnm{Regoli}, \binits{D.}},
\bauthor{\bsnm{Tardelli}, \binits{S.}},
\bauthor{\bsnm{Tesconi}, \binits{M.}}:
\bctitle{Fake: Evidence of spam and bot activity in stock microblogs on
  twitter}.
In: \bbtitle{Twelfth International AAAI Conference on Web and Social Media}
(\byear{2018})
\end{bchapter}
\endbibitem

%%% 17
\bibitem{mazza2019rtbust}
\begin{bchapter}
\bauthor{\bsnm{Mazza}, \binits{M.}},
\bauthor{\bsnm{Cresci}, \binits{S.}},
\bauthor{\bsnm{Avvenuti}, \binits{M.}},
\bauthor{\bsnm{Quattrociocchi}, \binits{W.}},
\bauthor{\bsnm{Tesconi}, \binits{M.}}:
\bctitle{Rtbust: Exploiting temporal patterns for botnet detection on twitter}.
In: \bbtitle{Proceedings of the 10th ACM Conference on Web Science},
pp. \bfpage{183}--\blpage{192}
(\byear{2019})
\end{bchapter}
\endbibitem

%%% 18
\bibitem{cai2017behavior}
\begin{bchapter}
\bauthor{\bsnm{Cai}, \binits{C.}},
\bauthor{\bsnm{Li}, \binits{L.}},
\bauthor{\bsnm{Zengi}, \binits{D.}}:
\bctitle{Behavior enhanced deep bot detection in social media}.
In: \bbtitle{2017 IEEE International Conference on Intelligence and Security
  Informatics (ISI)},
pp. \bfpage{128}--\blpage{130}
(\byear{2017}).
\bcomment{IEEE}
\end{bchapter}
\endbibitem

%%% 19
\bibitem{wu2021novel}
\begin{barticle}
\bauthor{\bsnm{Wu}, \binits{Y.}},
\bauthor{\bsnm{Fang}, \binits{Y.}},
\bauthor{\bsnm{Shang}, \binits{S.}},
\bauthor{\bsnm{Jin}, \binits{J.}},
\bauthor{\bsnm{Wei}, \binits{L.}},
\bauthor{\bsnm{Wang}, \binits{H.}}:
\batitle{A novel framework for detecting social bots with deep neural networks
  and active learning}.
\bjtitle{Knowledge-Based Systems}
\bvolume{211},
\bfpage{106525}
(\byear{2021})
\end{barticle}
\endbibitem

%%% 20
\bibitem{feng2021twibot}
\begin{bchapter}
\bauthor{\bsnm{Feng}, \binits{S.}},
\bauthor{\bsnm{Wan}, \binits{H.}},
\bauthor{\bsnm{Wang}, \binits{N.}},
\bauthor{\bsnm{Li}, \binits{J.}},
\bauthor{\bsnm{Luo}, \binits{M.}}:
\bctitle{Twibot-20: A comprehensive twitter bot detection benchmark}.
In: \bbtitle{Proceedings of the 30th ACM International Conference on
  Information \& Knowledge Management},
pp. \bfpage{4485}--\blpage{4494}
(\byear{2021})
\end{bchapter}
\endbibitem

%%% 21
\bibitem{al2018prediction}
\begin{barticle}
\bauthor{\bsnm{Al-Qurishi}, \binits{M.}},
\bauthor{\bsnm{Alrubaian}, \binits{M.}},
\bauthor{\bsnm{Rahman}, \binits{S.M.M.}},
\bauthor{\bsnm{Alamri}, \binits{A.}},
\bauthor{\bsnm{Hassan}, \binits{M.M.}}:
\batitle{A prediction system of sybil attack in social network using
  deep-regression model}.
\bjtitle{Future Generation Computer Systems}
\bvolume{87},
\bfpage{743}--\blpage{753}
(\byear{2018})
\end{barticle}
\endbibitem

%%% 22
\bibitem{chavoshi2016debot}
\begin{bchapter}
\bauthor{\bsnm{Chavoshi}, \binits{N.}},
\bauthor{\bsnm{Hamooni}, \binits{H.}},
\bauthor{\bsnm{Mueen}, \binits{A.}}:
\bctitle{Debot: Twitter bot detection via warped correlation.}
In: \bbtitle{Icdm},
vol. \bseriesno{18},
pp. \bfpage{28}--\blpage{65}
(\byear{2016})
\end{bchapter}
\endbibitem

%%% 23
\bibitem{minnich2017botwalk}
\begin{bchapter}
\bauthor{\bsnm{Minnich}, \binits{A.}},
\bauthor{\bsnm{Chavoshi}, \binits{N.}},
\bauthor{\bsnm{Koutra}, \binits{D.}},
\bauthor{\bsnm{Mueen}, \binits{A.}}:
\bctitle{Botwalk: Efficient adaptive exploration of twitter bot networks}.
In: \bbtitle{Proceedings of the 2017 IEEE/ACM International Conference on
  Advances in Social Networks Analysis and Mining 2017},
pp. \bfpage{467}--\blpage{474}
(\byear{2017})
\end{bchapter}
\endbibitem

%%% 24
\bibitem{mannocci2022mulbot}
\begin{bchapter}
\bauthor{\bsnm{Mannocci}, \binits{L.}},
\bauthor{\bsnm{Cresci}, \binits{S.}},
\bauthor{\bsnm{Monreale}, \binits{A.}},
\bauthor{\bsnm{Vakali}, \binits{A.}},
\bauthor{\bsnm{Tesconi}, \binits{M.}}:
\bctitle{Mulbot: Unsupervised bot detection based on multivariate time series}.
In: \bbtitle{2022 IEEE International Conference on Big Data (Big Data)},
pp. \bfpage{1485}--\blpage{1494}
(\byear{2022}).
\bcomment{IEEE}
\end{bchapter}
\endbibitem

%%% 25
\bibitem{sayyadiharikandeh2020detection}
\begin{bchapter}
\bauthor{\bsnm{Sayyadiharikandeh}, \binits{M.}},
\bauthor{\bsnm{Varol}, \binits{O.}},
\bauthor{\bsnm{Yang}, \binits{K.-C.}},
\bauthor{\bsnm{Flammini}, \binits{A.}},
\bauthor{\bsnm{Menczer}, \binits{F.}}:
\bctitle{Detection of novel social bots by ensembles of specialized
  classifiers}.
In: \bbtitle{Proceedings of the 29th ACM International Conference on
  Information \& Knowledge Management},
pp. \bfpage{2725}--\blpage{2732}
(\byear{2020})
\end{bchapter}
\endbibitem

%%% 26
\bibitem{dimitriadis2021social}
\begin{barticle}
\bauthor{\bsnm{Dimitriadis}, \binits{I.}},
\bauthor{\bsnm{Georgiou}, \binits{K.}},
\bauthor{\bsnm{Vakali}, \binits{A.}}:
\batitle{Social botomics: A systematic ensemble ml approach for explainable and
  multi-class bot detection}.
\bjtitle{Applied Sciences}
\bvolume{11}(\bissue{21}),
\bfpage{9857}
(\byear{2021})
\end{barticle}
\endbibitem

%%% 27
\bibitem{hurtado2019bot}
\begin{bchapter}
\bauthor{\bsnm{Hurtado}, \binits{S.}},
\bauthor{\bsnm{Ray}, \binits{P.}},
\bauthor{\bsnm{Marculescu}, \binits{R.}}:
\bctitle{Bot detection in reddit political discussion}.
In: \bbtitle{Proceedings of the Fourth International Workshop on Social
  Sensing},
pp. \bfpage{30}--\blpage{35}
(\byear{2019})
\end{bchapter}
\endbibitem

%%% 28
\bibitem{saeed2022trollmagnifier}
\begin{bchapter}
\bauthor{\bsnm{Saeed}, \binits{M.H.}},
\bauthor{\bsnm{Ali}, \binits{S.}},
\bauthor{\bsnm{Blackburn}, \binits{J.}},
\bauthor{\bsnm{De~Cristofaro}, \binits{E.}},
\bauthor{\bsnm{Zannettou}, \binits{S.}},
\bauthor{\bsnm{Stringhini}, \binits{G.}}:
\bctitle{Trollmagnifier: Detecting state-sponsored troll accounts on reddit}.
In: \bbtitle{2022 IEEE Symposium on Security and Privacy (SP)},
pp. \bfpage{2161}--\blpage{2175}
(\byear{2022}).
\bcomment{IEEE}
\end{bchapter}
\endbibitem

%%% 29
\bibitem{akyon2019instagram}
\begin{bchapter}
\bauthor{\bsnm{Akyon}, \binits{F.C.}},
\bauthor{\bsnm{Kalfaoglu}, \binits{M.E.}}:
\bctitle{Instagram fake and automated account detection}.
In: \bbtitle{2019 Innovations in Intelligent Systems and Applications
  Conference (ASYU)},
pp. \bfpage{1}--\blpage{7}
(\byear{2019}).
\bcomment{IEEE}
\end{bchapter}
\endbibitem

%%% 30
\bibitem{zarei2019typification}
\begin{bchapter}
\bauthor{\bsnm{Zarei}, \binits{K.}},
\bauthor{\bsnm{Farahbakhsh}, \binits{R.}},
\bauthor{\bsnm{Crespi}, \binits{N.}}:
\bctitle{Typification of impersonated accounts on instagram}.
In: \bbtitle{2019 IEEE 38th International Performance Computing and
  Communications Conference (IPCCC)},
pp. \bfpage{1}--\blpage{6}
(\byear{2019}).
\bcomment{IEEE}
\end{bchapter}
\endbibitem

%%% 31
\bibitem{beskow2019its}
\begin{barticle}
\bauthor{\bsnm{Beskow}, \binits{D.M.}},
\bauthor{\bsnm{Carley}, \binits{K.M.}}:
\batitle{Its all in a name: detecting and labeling bots by their name}.
\bjtitle{Computational and mathematical organization theory}
\bvolume{25}(\bissue{1}),
\bfpage{24}--\blpage{35}
(\byear{2019})
\end{barticle}
\endbibitem

%%% 32
\bibitem{uyheng2021active}
\begin{barticle}
\bauthor{\bsnm{Uyheng}, \binits{J.}},
\bauthor{\bsnm{Ng}, \binits{L.H.X.}},
\bauthor{\bsnm{Carley}, \binits{K.M.}}:
\batitle{Active, aggressive, but to little avail: characterizing bot activity
  during the 2020 singaporean elections}.
\bjtitle{Computational and Mathematical Organization Theory}
\bvolume{27}(\bissue{3}),
\bfpage{324}--\blpage{342}
(\byear{2021})
\end{barticle}
\endbibitem

%%% 33
\bibitem{Luceri_Deb_Giordano_Ferrara_2019}
\begin{botherref}
\oauthor{\bsnm{Luceri}, \binits{L.}},
\oauthor{\bsnm{Deb}, \binits{A.}},
\oauthor{\bsnm{Giordano}, \binits{S.}},
\oauthor{\bsnm{Ferrara}, \binits{E.}}:
Evolution of bot and human behavior during elections.
First Monday
\textbf{24}(9)
(2019).
\doiurl{10.5210/fm.v24i9.10213}
\end{botherref}
\endbibitem

%%% 34
\bibitem{Ferrara_2017}
\begin{botherref}
\oauthor{\bsnm{Ferrara}, \binits{E.}}:
Disinformation and social bot operations in the run up to the 2017 french
  presidential election.
First Monday
\textbf{22}(8)
(2017).
\doiurl{10.5210/fm.v22i8.8005}
\end{botherref}
\endbibitem

%%% 35
\bibitem{beskow2018bot}
\begin{bchapter}
\bauthor{\bsnm{Beskow}, \binits{D.M.}},
\bauthor{\bsnm{Carley}, \binits{K.M.}}:
\bctitle{Bot-hunter: a tiered approach to detecting \& characterizing automated
  activity on twitter}.
In: \bbtitle{Conference Paper. SBP-BRiMS: International Conference on Social
  Computing, Behavioral-Cultural Modeling and Prediction and Behavior
  Representation in Modeling and Simulation},
vol. \bseriesno{3},
p. \bfpage{3}
(\byear{2018})
\end{bchapter}
\endbibitem

%%% 36
\bibitem{githubGitHubMkearneytweetbotornot}
\begin{botherref}
\oauthor{\bsnm{Kearney}, \binits{M.W.}}:
{G}it{H}ub - mkearney/tweetbotornot: {R} package for detecting {T}witter bots
  via machine learning --- github.com.
\url{https://github.com/mkearney/Tweetbotornot}.
[Accessed 06-09-2023]
(2018)
\end{botherref}
\endbibitem

%%% 37
\bibitem{Livingstone_2022}
\begin{botherref}
\oauthor{\bsnm{Livingstone}, \binits{R.M.}}:
Trump bots and algorithmic experimentation on twitter.
First Monday
\textbf{27}(11)
(2022).
\doiurl{10.5210/fm.v27i11.12392}
\end{botherref}
\endbibitem

%%% 38
\bibitem{murdock2023identifying}
\begin{barticle}
\bauthor{\bsnm{Murdock}, \binits{I.}},
\bauthor{\bsnm{Carley}, \binits{K.M.}},
\bauthor{\bsnm{Ya{\u{g}}an}, \binits{O.}}:
\batitle{Identifying cross-platform user relationships in 2020 us election
  fraud and protest discussions}.
\bjtitle{Online Social Networks and Media}
\bvolume{33},
\bfpage{100245}
(\byear{2023})
\end{barticle}
\endbibitem

%%% 39
\bibitem{baumgartner2020pushshift}
\begin{bchapter}
\bauthor{\bsnm{Baumgartner}, \binits{J.}},
\bauthor{\bsnm{Zannettou}, \binits{S.}},
\bauthor{\bsnm{Keegan}, \binits{B.}},
\bauthor{\bsnm{Squire}, \binits{M.}},
\bauthor{\bsnm{Blackburn}, \binits{J.}}:
\bctitle{The pushshift reddit dataset}.
In: \bbtitle{Proceedings of the International AAAI Conference on Web and Social
  Media},
vol. \bseriesno{14},
pp. \bfpage{830}--\blpage{839}
(\byear{2020})
\end{bchapter}
\endbibitem

%%% 40
\bibitem{hayawi2022deeprobot}
\begin{barticle}
\bauthor{\bsnm{Hayawi}, \binits{K.}},
\bauthor{\bsnm{Mathew}, \binits{S.}},
\bauthor{\bsnm{Venugopal}, \binits{N.}},
\bauthor{\bsnm{Masud}, \binits{M.M.}},
\bauthor{\bsnm{Ho}, \binits{P.-H.}}:
\batitle{Deeprobot: a hybrid deep neural network model for social bot detection
  based on user profile data}.
\bjtitle{Social Network Analysis and Mining}
\bvolume{12}(\bissue{1}),
\bfpage{43}
(\byear{2022})
\end{barticle}
\endbibitem

%%% 41
\bibitem{10.3389/fdata.2023.1221744}
\begin{botherref}
\oauthor{\bsnm{Ng}, \binits{L.H.X.}},
\oauthor{\bsnm{Carley}, \binits{K.M.}}:
Do you hear the people sing? comparison of synchronized url and narrative
  themes in 2020 and 2023 french protests.
Frontiers in Big Data
\textbf{6}
(2023).
\doiurl{10.3389/fdata.2023.1221744}
\end{botherref}
\endbibitem

%%% 42
\bibitem{khaund2021social}
\begin{barticle}
\bauthor{\bsnm{Khaund}, \binits{T.}},
\bauthor{\bsnm{Kirdemir}, \binits{B.}},
\bauthor{\bsnm{Agarwal}, \binits{N.}},
\bauthor{\bsnm{Liu}, \binits{H.}},
\bauthor{\bsnm{Morstatter}, \binits{F.}}:
\batitle{Social bots and their coordination during online campaigns: A survey}.
\bjtitle{IEEE Transactions on Computational Social Systems}
\bvolume{9}(\bissue{2}),
\bfpage{530}--\blpage{545}
(\byear{2021})
\end{barticle}
\endbibitem

%%% 43
\bibitem{pacheco2021uncovering}
\begin{bchapter}
\bauthor{\bsnm{Pacheco}, \binits{D.}},
\bauthor{\bsnm{Hui}, \binits{P.-M.}},
\bauthor{\bsnm{Torres-Lugo}, \binits{C.}},
\bauthor{\bsnm{Truong}, \binits{B.T.}},
\bauthor{\bsnm{Flammini}, \binits{A.}},
\bauthor{\bsnm{Menczer}, \binits{F.}}:
\bctitle{Uncovering coordinated networks on social media: methods and case
  studies}.
In: \bbtitle{Proceedings of the International AAAI Conference on Web and Social
  Media},
vol. \bseriesno{15},
pp. \bfpage{455}--\blpage{466}
(\byear{2021})
\end{bchapter}
\endbibitem

%%% 44
\bibitem{cresci2020decade}
\begin{barticle}
\bauthor{\bsnm{Cresci}, \binits{S.}}:
\batitle{A decade of social bot detection}.
\bjtitle{Communications of the ACM}
\bvolume{63}(\bissue{10}),
\bfpage{72}--\blpage{83}
(\byear{2020})
\end{barticle}
\endbibitem

%%% 45
\bibitem{arin2023deep}
\begin{barticle}
\bauthor{\bsnm{Arin}, \binits{E.}},
\bauthor{\bsnm{Kutlu}, \binits{M.}}:
\batitle{Deep learning based social bot detection on twitter}.
\bjtitle{IEEE Transactions on Information Forensics and Security}
\bvolume{18},
\bfpage{1763}--\blpage{1772}
(\byear{2023})
\end{barticle}
\endbibitem

%%% 46
\bibitem{adel2022security}
\begin{bchapter}
\bauthor{\bsnm{Adel~Alipour}, \binits{S.}},
\bauthor{\bsnm{Orji}, \binits{R.}},
\bauthor{\bsnm{Zincir-Heywood}, \binits{N.}}:
\bctitle{Security of social networks: Lessons learned on twitter bot analysis
  in the literature}.
In: \bbtitle{Proceedings of the 17th International Conference on Availability,
  Reliability and Security},
pp. \bfpage{1}--\blpage{9}
(\byear{2022})
\end{bchapter}
\endbibitem

%%% 47
\bibitem{ng2023botbuster}
\begin{bchapter}
\bauthor{\bsnm{Ng}, \binits{L.H.X.}},
\bauthor{\bsnm{Carley}, \binits{K.M.}}:
\bctitle{Botbuster: Multi-platform bot detection using a mixture of experts}.
In: \bbtitle{Proceedings of the International AAAI Conference on Web and Social
  Media},
vol. \bseriesno{17},
pp. \bfpage{686}--\blpage{697}
(\byear{2023})
\end{bchapter}
\endbibitem

\end{thebibliography}

\end{document}